\documentclass[a4paper]{jpconf}
\usepackage{graphicx}

\newcommand{\be}{\begin{equation}}\newcommand{\ee}{\end{equation}}
\newcommand{\bea}{\begin{eqnarray}}\newcommand{\eea}{\end{eqnarray}}
\newcommand{\brr}{\begin{array}}\newcommand{\err}{\end{array}}
\newcommand{\bit}{\begin{itemize}}\newcommand{\eit}{\end{itemize}}
\newcommand{\ben}{\begin{enumerate}}\newcommand{\een}{\end{enumerate}}

\newcommand{\ba}{\begin{array}}
\newcommand{\ea}{\end{array}}

\def\lab{\label}\def\lan{\langle}
\def\lf{\left}

\def\non{\nonumber}\def\ran{\rangle}
\def\rar{\rightarrow}
\def\ri{\right}
\def\al{\alpha}\def\bt{\beta}
\def\de{\delta}\def\De{\Delta}
\def\te{\theta}
\def\si{\sigma}
\def\om{\omega}

\def\1{{_{1}}}\def\2{{_{2}}}

\newcommand{\ide}{1\hspace{-1mm}{\rm I}}

\def\noHe0{:\;\!\!\;\!\!:H_e(0):\;\!\!\;\!\!:}
\def\noHm0{:\;\!\!\;\!\!:H_\mu(0):\;\!\!\;\!\!:}
\def\nof{:\;\!\!\;\!\!:}

\def\lab{\label}
\def\lan{\langle}
\def\lf{\left}

\def\non{\nonumber}
\def\ran{\rangle}
\def\rar{\rightarrow}
\def\ri{\right}

\def\al{\alpha}\def\bt{\beta}
\def\de{\delta}\def\De{\Delta}
\def\te{\theta}

\def\si{\sigma}
\def\om{\omega}

\def\1{{_{1}}}\def\2{{_{2}}}
\def\nof{:\;\!\!\;\!\!:}
\def\wQ{Q}
\def\wwQ{Q}

\begin{document}
\title{On entanglement in neutrino mixing and oscillations}

\author{Massimo Blasone$^{\dag \, 1,2}$, Fabio Dell'Anno$^{1,2,3}$, Silvio De Siena$^{1,2,3}$ and Fabrizio Illuminati$^{1,2,3,4}$ }

\address{$^{1}$ Dipartimento di Matematica e Informatica,
Universit\`a degli Studi di Salerno, Via Ponte don Melillo,
I-84084 Fisciano (SA), Italy
\\
  $^{2}$ INFN Sezione di Napoli,
Gruppo collegato di Salerno, Baronissi (SA), Italy
\\
$^{3}$ CNR-INFM Coherentia, Napoli, and CNISM, Unit\`a di Salerno, Italy
\\
$^{4}$
ISI
Foundation for Scientific Interchange, Viale Settimio Severo 65,
I-10133 Torino, Italy
}

\ead{$^{\dag}$blasone@sa.infn.it}

\begin{abstract}
We report on recent results about entanglement in the context of
particle mixing and oscillations. We study in detail  single-particle entanglement
arising in two-flavor neutrino mixing. The analysis is performed first in the
context of Quantum Mechanics, and then for the case of Quantum Field
Theory.
\end{abstract}

\section{Introduction}

Entanglement has been widely investigated in a
number of physical systems, ranging from
 condensed matter to
atomic physics, and quantum optics \cite{Nielsen}.
Also in the context of  particle physics, the role of entanglement has been
considered, see for instance Refs.~\cite{ParticPhysEntang}.

In this paper, we consider the entanglement associated to neutrino mixing and oscillations.
A detailed study of such a topic has been performed recently in the context of quantum
mechanics  \cite{Noi1,Noi2}.
Here we review the main results of these studies in the simplest case of two flavors.
We show that these results suggest a simple extension of the analysis to the relativistic domain,
thus providing the physical basis and the mathematical tools for the quantification
of entanglement in quantum field theory.

The phenomenon of particle mixing,
associated with a mismatch between flavor and mass of the particle,
appears in several instances: quarks, neutrinos, and
the neutral $K$-meson system \cite{Cheng-Li,ParticleData}.
Particle mixing is at the basis of important effects
as neutrino oscillations and $CP$ violation \cite{Pontecorvo}.
Flavor mixing for the case of three generations is described by the
Pontecorvo-Maki-Nakagawa-Sakata (PMNS) in the lepton instance \cite{PMNS,Pont}.

In the following we consider only the simplest case of two flavors. In such a case,
the PMNS matrix reduces to the $2 \times 2$
rotation Pontecorvo matrix $\mathbf{U}(\theta)$,
\begin{equation}
\mathbf{U}(\theta) = \left( \begin{array}{cc}
  \cos\theta & \sin\theta \\
  -\sin\theta & \cos\theta
\end{array}
\right) \,,
\end{equation}
which connects the neutrino states with definite flavor
with those with definite masses:
\begin{equation}\label{flavstates}
|\underline{\nu}^{(f)}\rangle \,=\, \mathbf{U}(\theta)
\, |\underline{\nu}^{(m)}\rangle
\label{fermix3}
\end{equation}
where $|\underline{\nu}^{(f)}\rangle \,=\, \left(
|\nu_e\rangle,|\nu_\mu\rangle\right)^{T}$
and  $|\underline{\nu}^{(m)}\rangle \,=\, \left(
|\nu_1\rangle,|\nu_2\rangle \right)^{T}$.

From Eq.~(\ref{fermix3}), we see that each flavor state is given
by a superposition of mass eigenstates, i.e. $|\nu_{\alpha} \rangle =
U_{\alpha 1} |\nu_{1}\rangle + U_{\alpha 2} |\nu_{2}\rangle$.
Let us recall that both
$\{|\nu_{\alpha}\rangle\}$ and $\{|\nu_{i}\rangle\}$ are
orthonormal, i.e. $\langle \nu_{\alpha}|\nu_{\beta}\rangle =
\delta_{\alpha,\beta}$ and $\langle \nu_{i}|\nu_{j}\rangle =
\delta_{i,j}$.

We now  establish the
following correspondence with two-qubit states:
\begin{eqnarray}\label{massqubits}
|\nu_{1}\rangle \equiv |1\rangle_{1} |0\rangle_{2}  \equiv
|10\rangle, \quad
|\nu_{2}\rangle \equiv |0\rangle_{1} |1\rangle_{2}
 \equiv |01\rangle,
\end{eqnarray}
where $|\rangle_{i}$ denotes states in the Hilbert space for neutrinos
with mass $m_i$.
Thus, the occupation number allows to interpret the flavor
states as constituted by entangled superpositions of the mass
eigenstates. Quantum entanglement emerges as a direct consequence of
the superposition principle.
It is important to remark that the Fock space associated with the
neutrino mass eigenstates is physically well defined.
In fact, at least in principle, the mass eigenstates can be produced
or detected in experiments performing extremely precise kinematical
measurements \cite{Kayser}.
In this framework, as discussed in Ref.~\cite{Noi1},
the quantum mechanical state (\ref{fermix3}) is entangled in the field modes,
although being a single-particle state.

Mode entanglement defined for single-photon states of the radiation field
or associated with systems of identical particles
has been discussed in Ref.~\cite{Zanardi}.
The concept of mode entanglement in single-particle
states has been widely discussed and is by now well established
\cite{Zanardi,singpart}. Successful experimental
realizations using single-photon states
have been reported as well \cite{singpartexp}.
Moreover, remarkably, the nonlocality of single-photon states has been experimentally
demonstrated \cite{singpartexp2}, verifying a theoretical prediction \cite{TanHardy}.
Furthermore, the existing schemes to probe nonlocality
in single-particle states have been generalized to
include massive particles of arbitrary type \cite{Vedral}.

In the dynamical regime, flavor mixing (and neutrino mass
differences) generates the phenomenon of neutrino oscillations. The
mass eigenstates $|\nu_j\rangle$ have definite masses $m_{j}$ and
definite energies $\om_{j}$.  Their propagation can be described by
plane wave solutions of the form $|\nu_j(t)\rangle = e^{-i \om_{j}t}
|\nu_{j}\rangle$. The time evolution of the flavor neutrino states
Eq.(\ref{flavstates}) is given by:
\begin{equation}
|\underline{\nu}^{(f)}(t)\rangle  = \mathbf{\widetilde{U}}(t)
|\underline{\nu}^{(f)}\rangle \,, \qquad
\mathbf{\widetilde{U}}(t) \equiv
\mathbf{U}(\theta) \, \mathbf{U}_{0}(t) \,
\mathbf{U}(\theta)^{-1} \,,
\label{flavstateevolution}
\end{equation}
where $|\underline{\nu}^{(f)}\rangle$ are the flavor states at
$t=0$, $\mathbf{U}_{0}(t) = diag (e^{-i \om_{1}t},e^{-i \om_{2}t})$,
and $\mathbf{\widetilde{U}}(t=0)=\ide $.

At time $t$, the probability associated with the transition
$\nu_{\alpha}\rightarrow\nu_{\beta}$ is
\begin{equation}
P_{\nu_{\alpha}\rightarrow\nu_{\beta}}(t) \,=\,
|\langle\nu_{\beta}|\nu_{\alpha}(t)\rangle|^{2}
\,=\,|\mathbf{\widetilde{U}}_{\alpha \beta}(t)|^{2} \, ,
\label{neutrinooscillation}
\end{equation}
where $\alpha,\beta = e, \mu\,.$
The explicit form for the transition probabilities
 in the two flavor case is:
\bea\label{oscillQM1}
P_{\nu_e\rightarrow\nu_e}(t) &=&
1-\sin ^{2}(2\theta )\sin
^{2}\left( \frac{\omega _{2}-\omega _{1}}{2}t\right),
\\ \label{oscillQM2}
P_{\nu_e\rightarrow\nu_\mu}(t) &=&
\sin ^{2}(2\theta )\sin
^{2}\left( \frac{\omega _{2}-\omega _{1}}{2}t\right).
\eea

Flavor neutrino states are well defined
in the context of Quantum Field Theory (QFT), where they are
obtained as eigenstates of the flavor neutrino charges \cite{BV95,BHV99}.
In the relativistic limit, the exact QFT flavor
states reduce to the usual Pontecorvo flavor states Eq.(\ref{flavstates}):
flavor modes are thus legitimate and physically well-defined
individual entities and mode entanglement can be defined and
studied in analogy with the static case of Ref.\cite{Noi1}.
We  can thus establish the following
correspondence with two-qubit states:
\begin{eqnarray}\label{flavorqubits}
 |\nu_{e}\rangle \equiv |1\rangle_{e} |0\rangle_{\mu} ,
 \quad
|\nu_{\mu}\rangle \equiv |0\rangle_{e} |1\rangle_{\mu}.
\end{eqnarray}
States $|0\rangle_{\alpha}$ and $|1\rangle_{\alpha}$
correspond, respectively, to the absence and the presence of a
neutrino in mode $\alpha$. Entanglement is thus established among
flavor modes, in a single-particle setting.
Eq.~(\ref{flavstateevolution}) can then be recast as
\begin{equation}
|\nu_{\alpha}(t)\rangle = \mathbf{\widetilde{U}}_{\alpha e}(t)
|1\rangle_{e} |0\rangle_{\mu}  +
\mathbf{\widetilde{U}}_{\alpha \mu}(t) |0\rangle_{e} |1\rangle_{\mu}
  \, ,
\label{flavorWstate}
\end{equation}
with the normalization condition
$\sum_{\beta}|\mathbf{\widetilde{U}}_{\alpha \beta}(t)|^{2}=1$
$(\alpha,\beta=e,\mu)$. The time-evolved states
$|\underline{\nu}^{(f)}(t)\rangle$ are entangled superpositions of
the
two flavor eigenstates with time-dependent coefficients.
Thus, flavor oscillations can be related to bipartite (flavor)
entanglement of single-particle states \cite{Noi2}.

\section{Entanglement in neutrino mixing -- Quantum Mechanics}
\label{Entmeasures}

As discussed in the Introduction, the flavor neutrino state at a given time,
say $|\nu_e(t)\ran$  for
definiteness, can be regarded as an entangled state either in terms of the mass eigenstates or
in terms of the flavor eigenstates (at a fixed time). In the first instance, which was studied in
detail for the multipartite case in Ref.\cite{Noi1}, we have a static entanglement,
in the sense that the result of the entanglement measures on the state
$|\nu_e(t)\ran$  do not depend on time.
In the second case, considered for the general three flavor case in Ref.\cite{Noi2},
the entanglement varies with time as it is related to the oscillations of flavor(s).

In this Section, we discuss these two forms of entanglement in the simple case of two flavors, by means of entropy measures first, and then using the characterization of entanglement in terms
of quantum uncertainties. The second approach turns out
to be suitable for the generalization of our
discussion to the case of mixing among relativistic quantum fields -- see next Section.

\subsection{Neutrino entanglement via linear entropy}

In terms of the mass eigenstates, the electron neutrino state at time $t$ reads:
\bea
|\nu_e(t)\rangle &=& e^{-i\om_1 t}\, \cos\te\,
|\nu_1\rangle\,  + \,e^{-i\om_2 t}\, \sin\te\,
|\nu_2\rangle  \, ,
\label{nuetm}
\eea
where $|\nu_i\rangle  $ are interpreted as the qubits, see Eq.(\ref{massqubits}).

Following the usual procedure, we construct the density operator $\rho^{(\alpha)}=|\nu_{\alpha}(t)\rangle\langle \nu_{\alpha}(t)|$
corresponding to the pure state $|\nu_{\alpha}(t)\rangle$. Then we consider the density
matrix
 $\rho_i^{(\alpha)} \,=\,Tr_{j}[\rho^{(\alpha)}]$
reduced with respect to the index $j$.  For the specific case of the state Eq.(\ref{nuetm}),
we have $\rho^{(e)}=|\nu_{e}(t)\rangle\langle \nu_{e}(t)|$ and
\bea\label{rhoem1}
 \rho_1^{(e)} \,=\,Tr_{2}[\rho^{(e)}]= \cos^2\te \, |1\ran_1\, {}_1\lan 1| \,+\,
 \sin^2\te \, |0\ran_1 \,{}_1\lan 0|
 \\ \label{rhoem2}
 \rho_2^{(e)} \,=\,Tr_{1}[\rho^{(e)}]= \cos^2\te \, |0\ran_2 \,{}_2\lan 0| \,+\,
 \sin^2\te \, |1\ran_2 \,{}_2\lan 1|
\eea
It is then easy to calculate the corresponding linear entropies, which turn out to be equal:
\bea
\label{lentem1}
S_{L}^{(1;2)}(\rho_e) \,=\, 2\lf(1-Tr_1[(\rho_1^{(e)})^{2}]\ri)\, = \,\sin^2(2\te) ,
\\ \label{lentem2}
S_{L}^{(2;1)}(\rho_e) \,=\, 2\lf(1-Tr_2[(\rho_2^{(e)})^{2}]\ri)\, = \,  \sin^2(2\te) ,
\eea

Similar results are found for the muon neutrino state. Note that the above results are particular
cases of the more general ones obtained for the three flavor neutrino states in Ref.\cite{Noi1},
where it was found that such states can be classified as generalized W states. In the present
(two-flavor) case, the form of the entangled state is simply that of a Bell state.

Eqs.(\ref{lentem1})-(\ref{lentem2}) express the fact that flavor neutrino states at any time can be regarded as entangled superpositions of the mass qubits $|\nu_i\ran$, where the entanglement is a function of
the mixing angle only.

\smallskip

Let us now turn to the dynamic entanglement arising in connection with flavor oscillations.
To this aim, we rewrite the electron neutrino state $|\nu_{e}(t)\rangle$ as
\bea
|\nu_e(t)\rangle &=& \mathbf{\widetilde{U}}_{e e}(t)\,
|\nu_e\rangle\,  + \,\mathbf{\widetilde{U}}_{e \mu}(t)
|\nu_\mu\rangle  \, ,
\label{nuetm2}
\eea
where $|\nu_e\rangle $, $|\nu_\mu\rangle$ are the flavor neutrino states at time $t=0$ and
are now taken as the relevant qubits  (see
Eq.(\ref{flavorqubits})).
By proceeding in a similar way as done for the static case, we
arrive at the following expression for the linear entropies associated to the above state:
\begin{eqnarray}
S_{L}^{(\mu;e)}(\rho_e) \,=\, S_{L}^{(e;\mu)}(\rho_e)
&=& 4 |\mathbf{\widetilde{U}}_{e e}(t)|^{2}
\, |\mathbf{\widetilde{U}}_{e \mu}(t)|^{2}
\nonumber \\ \label{SLsinglet}
&=& 4
|\mathbf{\widetilde{U}}_{e e}(t)|^{2} \,
(1-|\mathbf{\widetilde{U}}_{e e}(t)|^{2})
\end{eqnarray}
 Eq.(\ref{SLsinglet}) establishes that the linear entropy of the reduced
state is equal to the product of the two-flavor transition
probabilities given in Eqs.(\ref{neutrinooscillation})-(\ref{oscillQM2}).
It is remarkable that simple expressions  similar to those of
Eq.~(\ref{SLsinglet})  hold also for the three flavor case \cite{Noi2}.

Note also that, for any reduced state $\rho$ of a two-level
system one has that $S_L = 2[1-Tr(\rho^{2})] = 4Det\rho =
4\lambda_1(1-\lambda_1)$, where $\lambda_1$ is one of the two
non-negative eigenvalues of $\rho$, and the relation $\lambda_1 +
\lambda_2 = 1$ has been exploited. Comparing with
Eq.~(\ref{SLsinglet}), one sees that the transition probabilities
coincide
with the eigenvalues of the reduced state density matrix.

\begin{figure}[t]
\centering
\includegraphics*[width=7.5cm]{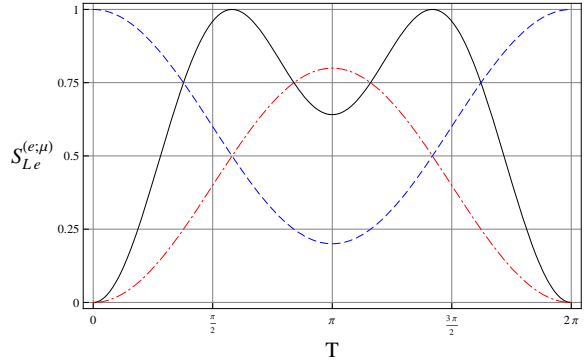}
\caption{(Color online) Linear entropy $S_{L}^{(e;\mu)}(\rho_e)$ (full) as a
function of the scaled time $T=\frac{2Et}{\Delta m_{12}^{2}}$.
The mixing angle $\theta$ is fixed at the experimental
value $\sin^{2}\theta = 0.314$.
The transition probabilities
$P_{\nu_{e}\rightarrow\nu_{e}}$ (dashed) and
$P_{\nu_{e}\rightarrow\nu_{\mu}}$ (dot-dashed) are
reported as well for comparison.}
\label{FigTwoFlav}
\end{figure}

In Fig.~\ref{FigTwoFlav} we show the behavior of$ S_{L}^{(e;\mu)}(\rho_e)$
as a function of the scaled, dimensionless time
$T=\frac{2 E t}{\Delta m_{12}^{2}}$. In the same figure, we also
report the behavior of the transition probabilities
$P_{\nu_{e}\rightarrow\nu_{e}}$ and
$P_{\nu_{e}\rightarrow\nu_{\mu}}$.
The plots have a clear physical interpretation. At time $T=0$, the
entanglement is zero, the global state of the system is factorized,
and the two flavors are not mixed. For $T>0$, flavors start to
oscillate and the entanglement is maximal at largest mixing:
$P_{\nu_{e}\rightarrow\nu_{e}}=P_{\nu_{e}\rightarrow\nu_{\mu}}=0.5$,
and minimum  at $T=\pi$.

\subsection{Neutrino entanglement via uncertainties}

An alternative characterization of entanglement can be given
in terms of quantum uncertainties \cite{klyachko1,klyachko2}. In order to apply this
formalism to the case of neutrino
mixing and oscillations we introduce the (fermionic)
annihilation  operator $\al_i $ for a neutrino with mass $m_i$, with anti-commutator
$\{\al_i,\al_j\}=\de_{ij}$.
We then  define neutrino states with definite masses as:
\bea
|\nu_i\ran \equiv \al^\dag_i |0\ran_m\,,\quad i=1,2
\eea
where  $|0\ran_m\equiv |0\ran_1\otimes|0\ran_2$ is the vacuum for the mass eigenstates.

Next we define the flavor annihilation operators  by means of the
following relations:
\bea
\al_e(t) &=& \cos\te \, \al_1(t) \,+\,  \sin\te \, \al_2(t)
\\ [2mm]
\al_\mu(t) &=& -\sin\te \, \al_2(t) \,+\,  \cos\te \, \al_1(t)
\eea
where  $\al_i(t)=e^{i \om_i t} \al_i$, with $i=1,2$.

The flavor states are given by:
\bea \label{flavstateQM}
|\nu_\si (t)\ran \equiv \al_\si^\dag (t)|0\ran_m, \quad \si=e,\mu.
\eea
We use in the following the notation $|\nu_\si \ran\equiv |\nu_\si (t=0)\ran$.

The hamiltonian for the system is
\bea \non
H&=& \om_{ee} \al^\dag_e(t)\al_e(t) \, +\,\om_{\mu\mu} \al^\dag_\mu(t)\al_\mu(t) \, +\,
\om_{e\mu} \lf(\al^\dag_e(t)\al_\mu(t) \, +\,\al^\dag_\mu(t)\al_e(t) \ri)
\\
&=& \om_1 \al^\dag_1\al_1 \, +\,\om_2 \al^\dag_2\al_2
\eea
where we used the relations $ \om_{ee} =
\om_{1}\cos^{2}\te + \om_{2} \sin^{2}\te~$,  $\om_{\mu\mu} =
\om_{1}\sin^{2}\te + \om_{2} \cos^{2}\te~$,   $\om_{e\mu}
=(\om_{2}-\om_{1})\sin\te \cos\te\,$.

In the mass basis, we can introduce the $SU(2)$ operators (the superscript
$m$ stands for mass):
\bea
J_+^m &=& \al_1^\dag \al_2
\quad , \qquad
J_-^m \,=\, \al_2^\dag \al_1,
\\
J_3^m &=& \frac{1}{2}\lf(\al_1^\dag \al_1 \, -\,\al_2^\dag \al_2\ri)
\\
{\cal C} &=&  \frac{1}{2}\lf(\al_1^\dag \al_1 \, +\,\al_2^\dag \al_2\ri)
\eea
We also have
\bea
N_1 &=& {\cal C} \,+\, J_3^m
\\
N_2 &=& {\cal C} \,-\, J_3^m
\\
H&=& \om_1 N_1\, +\, \om_2 N_2
\eea

The static entanglement of the electron  neutrino state $|\nu_e(t)\ran$ defined in Eq.(\ref{flavstateQM}), is characterized, in the present formalism, by the variances associated with
the numbers $N_i$, relative to the mass qubits.
Thus we have:
\bea \non
\De N_i(\nu_e) &\equiv& \lan \nu_e(t)| N_i^2| \nu_e(t)\ran \,-\,   \lan \nu_e(t)| N_i| \nu_e(t)\ran^2
\\ \label{statentQM}
&=& \frac{1}{4} \sin^2(2\te)\,, \qquad i=1,2.
\eea
This result differs by a factor 4 from that obtained by means of the linear entropy, Eqs.(\ref{lentem1})-(\ref{lentem2}).

\smallskip

In order to discuss the dynamical entanglement of the state $|\nu_e(t)\ran$, we need to
introduce flavor oscillations, which
can be seen both in terms of overlaps of states at different times:
\bea
P_{\nu_e\rar\nu_e}(t)&=& |\lan \nu_e|\nu_e(t)\ran|^2
\\
P_{\nu_e\rar\nu_\mu}(t)&=& |\lan \nu_\mu|\nu_e(t)\ran|^2
\eea
with $P_{\nu_e\rar\nu_e}(t)+P_{\nu_e\rar\nu_\mu}(t)=1$,
or equivalently in terms of expectation values of number operators at time $t$:
\bea
P_{\nu_e\rar\nu_e}(t)&=& \lan \nu_e|N_e(t)|\nu_e\ran
\\
P_{\nu_e\rar\nu_\mu}(t)&=& \lan \nu_e|N_\mu(t)|\nu_e\ran
\\
N_\si(t)&=&\al^\dag_\si(t)\al_\si(t) \quad \si=e,\mu
\eea
The explicit expressions of the transition probabilities
are given in Eqs.(\ref{oscillQM1}),(\ref{oscillQM2}).

In the flavor basis, we can again introduce  $SU(2)$ operators (at time $t$):
\bea
J_+^f(t) &=& \al_e^\dag(t)\al_\mu(t)
\quad, \quad
J_-^f(t) \,=\, \al_\mu^\dag(t) \al_e(t)
\\
J_3^f(t) &=& \frac{1}{2}\lf(\al_e^\dag(t) \al_e(t) \, -\,\al_\mu^\dag(t) \al_\mu(t)\ri)
\\
{\cal C} &=& \frac{1}{2}\lf(\al_e^\dag(t) \al_e(t) \, +\,\al_\mu^\dag(t) \al_\mu(t)\ri)
\,=\, \frac{1}{2}\lf(\al_1^\dag \al_1 \, +\,\al_2^\dag \al_2\ri)
\eea
where the superscript $f$ stands for flavor. The flavor number operators are:
\bea
N_e(t) &=& {\cal C} \,+\, J_3^f(t)
\\
N_\mu(t) &=& {\cal C} \,-\, J_3^f(t)
\eea
Note that the Hamiltonian can be written as:
\bea
H&=& \om_{ee} N_e(t) \, +\,\om_{\mu\mu} N_\mu(t) \, +\,
\om_{e\mu} (J_+^f (t)\, +\,J_-^f(t) )
\eea

Flavor entanglement is given by the variances of the above flavor numbers. Consider first
the the variances of
the $SU(2)$ operators in the flavor basis, which give:
\bea
\De J_1(\nu_e)(t) &=& \frac{1}{4} \left[1-\sin^2(4 \theta)
\sin^4\left(\frac{\omega_2-\omega_1}{2} t \right)\right]
\\
\De J_2(\nu_e)(t) &=&
\frac{1}{4}-  \sin^2(2\theta) \sin^2[(\omega_2-\omega_1) t]
\\
\De J_3(\nu_e)(t) &=&\sin^2(2 \theta)
\sin^2\left(\frac{\omega_2-\omega_1}{2} t \right)
\lf[1-  \sin^2(2 \theta)
\sin^2\left(\frac{\omega_2-\omega_1}{2} t \right)\ri]
\\
\De {\cal C}(\nu_e)&=& 0
\eea

When we restrict to a given flavor (e.g. electron neutrinos),
we find that the entanglement is
given by
\bea \label{DeNQM}
\De N_e(\nu_e)(t) &\equiv& \lan \nu_e(t)| N_e^2(t)| \nu_e(t)\ran \,-\,   \lan \nu_e(t)| N_e(t)| \nu_e(t)\ran^2
\\ [2mm]
 &=& P_{\nu_e\rar\nu_e}(t) \, P_{\nu_e\rar \nu_\mu}(t)
\eea
with the same result for $\De N_\mu(\nu_e)(t)$. The above result coincides (again up to a factor 4)
with the one obtained in
Eq.(\ref{SLsinglet}) by means of
 the
linear entropy.

Analogous results are easily obtained for the state $| \nu_\mu(t)\ran $.

\section{Neutrino mixing in Quantum Field Theory}

We now look for an extension of the above discussion to a relativistic context. To this aim,
we need to consider neutrino mixing at level of  fields \cite{BV95}.
For two flavors, the mixing transformations are
 \bea \nu _{e}(x) &=&\cos \theta \,\nu_1(x) \,+ \,\sin \theta \,\nu _{2}(x)
\\[2mm]
 \label{fieldmix} \nu _{\mu }(x)
&=&-\sin \theta\, \nu _{1}(x)\, +\,\cos \theta \,\nu _{2}(x)\, ,
\eea
where
$\nu_{e}(x)$ and $\nu_{\mu}(x)$ are the Dirac neutrino fields with
definite flavors. Here, $\nu_{1}(x)$ and $\nu_{2}(x)$ are the free
 neutrino
fields with definite masses $m_{1}$ and $m_{2}$, respectively. For the purpose of discussing
the quantization of flavor fields and the phenomenon of neutrino oscillations, it is sufficient
to consider only the free Lagrangian term for these two free  fields:
\bea\label{lagrmas}
{\cal L}_{\nu}(x)\,=\,  {\bar \nu_m}(x) \lf( i \not\!\partial -
M^{d}_{\nu} \ri) \nu_m(x)\, ,
\eea
where $\nu_m^T=(\nu_1,\nu_2)$ and $M^{d}_{\nu} = diag(m_{1},m_{2})$.
${\cal L}_{\nu}(x)$ is invariant
under global $U(1)$ phase transformations of the type $ \nu
_{m}^{^{\prime}}(x)=e^{i\alpha}\nu_{m}(x).$
This implies the conservation
of the Noether charge $ Q =\int I^{0}(x)d^{3}{\bf x}$ (with
$I^{\mu}(x)=\bar{\nu}_{m}(x)\gamma ^{\mu }\nu_{m}(x)$) which is
indeed the total charge of the system, i.e. the total lepton
number of neutrinos.

Consider now the global $SU(2)$ transformation \cite{BJV01}:
\bea\label{SU(2)tran}
 \nu_{m}^{ \prime }(x)=e^{i\alpha_{j}\cdot \tau_{j}}\nu
_{m}(x) \qquad \qquad  j=1,2,3.
 \eea
 with $\alpha_{j}$ real
constants, ${\tau_{j}} =\sigma_{j}/2$ with $\sigma_{j}$ being the
Pauli matrices.

${\cal L}_{\nu}$
is not invariant under the transformations (\ref{SU(2)tran}) since $m_{1} \neq m_{2}$. By use of the
equations of motion, we obtain
\bea
 \delta {\cal L}_{\nu} = i \alpha_{j}\bar{\nu}_{m}(x)\left[\tau_{j}
,\;M^{d}_{\nu} \right] \nu_{m}(x)=-\alpha_{j}\partial _{\mu}J_{m,j}^{\mu}(x)\,,
 \eea
 where the currents are:
  \bea
   J_{m,j}^{\mu
}(x)=\bar{\nu}_{m}(x)\;\gamma ^{\mu }\;\tau _{j}\;\nu
_{m}(x)\,,\quad\;\quad\;\quad j=1,2,3.
 \eea

The related charges
\bea\label{Qmj}
Q_{m,j}(t)=\int d^{3}{\bf x}\,
J_{m,j}^{0}(x)\,,
\eea
satisfy the $su(2)$ algebra:
\bea
&&\left[
Q_{m,i}(t),Q_{m,j}(t)\right] =i\varepsilon _{ijk}Q_{m,k}(t).
\eea
The Casimir operator is proportional to the total
(conserved) charge: $ Q_{m,0}=\frac{1}{2}Q_{\nu}$
and also $Q_{m,3}$ is conserved, due to the fact that
 $M^{d}_{\nu}$ is diagonal. This implies that the  charges for $\nu _{1}$ and $\nu _{2}$
 are separately  conserved.
The $U(1)$ Noether charges associated with  $\nu_1$ and $\nu_2$ can be  then expressed as
\bea\label{su2noether}
&&Q_1\, \equiv \,\frac{1}{2}Q \,+ \,Q_{m,3}\,;
\qquad \qquad Q_2\, \equiv \,\frac{1}{2}Q_{\nu} \,- \,Q_{m,3}\,.
\\
&&Q_i\, = \,\int d^{3}{\bf x} \,  \nu_{i}^{\dag}(x)\;\nu_{i}(x)\,,
\eea
with $Q$ total (conserved) charge and $i=1,2$.

\vspace{3mm}

Let us now consider the Lagrangian ${\cal L}_{\nu}(x)$ written in the flavor basis
\bea\label{lagrflav}
{\cal
L}_{\nu}(x)\,=\,  {\bar \nu_f}(x) \lf( i \not\!\partial - M_{\nu}
\ri) \nu_f(x)\, ,
\eea
where $\nu_{f}^{T}=(\nu _{e},$ $\nu
_{\mu })\,$ and $ M_{\nu}\,=\,  \lf(\ba{cc}m_{\nu_e} & m_{\nu_{e\mu}}
\\ m_{\nu_{e\mu}} & m_{\nu_\mu}\ea\ri)   $.

The variation of the
Lagrangian (\ref{lagrflav}) under the $SU(2)$ transformation:
\bea
\nu _{f}^{\prime }(x)=e^{i\alpha_{j}\cdot \tau_{j}}\nu
_{f}(x) \qquad \qquad  j=1,2,3\,,
 \eea
is given by
\bea
\delta {\cal L}_{\nu}(x)=i\alpha_{j}\bar{\nu}_{f}(x)\left[
\tau_{j}, M_{\nu}\right] \nu_{f}(x)=
-\alpha_{j}\partial _{\mu
}J_{f,j}^{\mu }(x)\,,
\eea
where
\bea J_{f,j}^{\mu
}(x)=\bar{\nu}_{f}(x)\;\gamma ^{\mu }\;\tau_{j}\; \nu _{f}(x)\,,
\quad \;\quad j=1,2,3\,.\eea
Again, the charges
\bea
Q_{f,j}(t)=\int d^{3}{\bf x} \,J_{f,j}^{0}(x)
\eea
close the $su(2)$ algebra, however, because of the off-diagonal (mixing) terms in
 $M_{\nu}$, $Q_{f,3}(t)$ is  time dependent. This
implies an exchange of charge between $\nu_{e}$ and $\nu_{\mu }$,
resulting in the phenomenon of neutrino oscillations.
 The (time dependent) flavor charges for mixed fields are then defined as \cite{BJV01}:
\bea\label{flavch} && Q_{e}(t)=\frac{1}{2}Q + Q_{f,3}(t)~,
\qquad Q_{\mu}(t)=\frac{1}{2}Q - Q_{f,3}(t)~,
 \\ \label{flavcharges}
&& Q_{\sigma}(t) \,= \,\int d^{3}{\bf x}\,
 \nu_{\sigma}^{\dag}(x)\;\nu_{\sigma}(x)~,
\eea
where $\sigma=e,\mu$ and
$Q_{e}(t) + Q_{\mu}(t) =
Q\,.$

\subsection{Flavor states for mixed neutrinos \label{S3}}

Till now our considerations have been essentially classical.
We now quantize the fields with
definite masses as usual (see Appendix), and consider the eigenstates of the
above defined charges.

The  normal ordered charge operators for free neutrinos $\nu_{1}$,
$\nu_{2}$ are:
\bea
:Q_{i}:\, \equiv \,\int d^{3}{\bf x} \, : \nu_{i}^{\dag}(x)\;\nu_{i}(x):
 = \, \sum_{r}
\int d^3 {\bf k} \, \lf( \al^{r\dag}_{{\bf k},i}
\al^{r}_{{\bf k},i}\, -\, \beta^{r\dag}_{-{\bf
k},i}\beta^{r}_{-{\bf k},i}\ri),
\eea
where $i=1,2$ and $:..:$ denotes normal ordering with respect to the vacuum $|0\ran_{1,2}$.
The neutrino states with definite masses  defined as
\bea
|\nu^{r}_{\;{\bf k},i}\ran =
\al^{r\dag}_{{\bf k},i} |0\ran_{1,2}, \qquad i=1,2,
\eea
are clearly eigenstates of $Q_{1}$ and $Q_{2}$, which can be identified with the
lepton charges of neutrinos in the absence of mixing.

The situation is more delicate when mixing is present. In such a
case, the flavor neutrino states have to be defined as the
eigenstates of the flavor charges $Q_{\sigma}(t)$ (at a given
time). The relation between the flavor charges in the presence of
mixing and those in the absence of mixing is:
\bea\label{carichemix1}
Q_{e}(t) &=& \cos^2\te\;  Q_{1} + \sin^2\te \; Q_{2}
+ \sin\te\cos\te \int d^3{\bf x} \lf[\nu_1^\dag (x) \nu_2(x) + \nu_2^\dag(x) \nu_1(x)\ri]\,,
\\
\label{carichemix2}
Q_{\mu}(t)&=& \sin^{2}\te \; Q_{1}  +\cos^{2}\te \; Q_{2}
- \sin\te \cos\te \int d^3{\bf x} \lf[\nu_1^\dag(x) \nu_2(x) + \nu_2^\dag(x) \nu_1(x)\ri]\,.
\eea
Notice that the last term in these expressions is proportional to
the charge  $Q_{m,1}$ defined above (cf. Eq.(\ref{Qmj})). The
presence of such a term forbids the construction of eigenstates of
the $Q_{\sigma}(t)$ in the Hilbert space ${\cal H}_{1,2}$.
One then is led \cite{BV95}  to define another Hilbert space,  ${\cal H}_{e,\mu}$
for the flavor neutrino fields. The flavor vacuum state $|0\ran_{e,\mu}$ and the
mass vacuum state
$|0\ran_{1,2}$ are orthogonal to each other (see Appendix).

The normal ordered flavor charge operators for mixed neutrinos are then written as
\bea\non
\nof \wwQ_{\sigma}(t) \nof\, &\equiv& \,\int d^{3}{\bf x}\,
\nof \nu_{\sigma}^{\dag}(x)\;\nu_{\sigma}(x) \nof \
\\ &=& \, \sum_{r}
\int d^3 {\bf k} \, \lf( \al^{r\dag}_{{\bf k},\si}(t)
\al^{r}_{{\bf k},\si}(t)\, -\, \beta^{r\dag}_{-{\bf
k},\si}(t) \beta^{r}_{-{\bf k},\si}(t)\ri)\,
\eea
where $\sigma=e,\mu$, and $\nof ... \nof\,$
denotes normal ordering with respect to  $|0\ran_{e,\mu}$.
Thus, the flavor charges are diagonal in the flavor annihilation/creation operators constructed
by means of the mixing generator $G_\theta$ defined in the Appendix.
The definition of the normal ordering  $\nof ... \nof\,$ for any operator $A$,
is the usual one:
\bea \lab{nordf}
 \nof A \nof\, \equiv \,  A \, -\, {}_{e,\mu}\lan 0| A | 0 \ran_{e,\mu}\,.
\eea
Note that
$
 \nof \wwQ_{\sigma}(t) \nof \; = \,G_\theta^{-1}(t)  :\wQ_{j} :
G_\theta(t),
$
 with $(\sigma,j) = (e,1),(\mu,2),$ and
\bea
\nof {Q} \nof \; = \;
 \nof {\wwQ}_{e}(t) \nof \;+\; \nof{\wwQ}_{\mu}(t)\nof
\;= \; :{\wQ}_{1}:\; +\; :\wQ_{2}: \; = \; :\wQ:\,.
\eea

We define the  flavor states as eigenstates of the flavor
charges $\wwQ_{\sigma} $ at a reference  time $t=0$:
\bea\label{flavstate}
|\nu^{r}_{\;{\bf k},\si}\ran \equiv \al^{r\dag}_{{\bf k},\sigma}(0) |0(0)\ran_{{e,\mu}},
\qquad \si = e,\mu\eea
 and similar ones for antiparticles.
We have
\bea \label{su2flavstates}
\nof \wwQ_{e}(0)\nof|\nu^{r}_{\;{\bf k},e}\rangle\,=\,|\nu^{r}_{\;{\bf k},e}\rangle \,,\qquad \qquad
\nof\wwQ_{\mu}(0)\nof|\nu^{r}_{\;{\bf k},\mu}\rangle\,=\,|\nu^{r}_{\;{\bf k},\mu}\rangle
\eea
\bea
\nof\wwQ_{e}(0)\nof\,|\nu^{r}_{\;{\bf k},\mu}\rangle\,=\,0\,=\;
\nof\wwQ_{\mu}(0)\nof\,|\nu^{r}_{\;{\bf k},e}\rangle\,,
\;\;\quad
 \nof\wwQ_{\sigma}(0)\nof\;|0\rangle_{{e,\mu}}\,=\,0.
 \eea
The explicit form of the flavor
states $|\nu_{{\bf k},e}^{r} \rangle $
 and $|\nu_{{\bf k},\mu}^{r}  \rangle $ at time $t=0$
 is given in the Appendix.

\subsection{Oscillation formulas in QFT}

Flavor oscillation formulas are derived
by computing, in the Heisenberg representation, the
expectation value of the flavor charge operators on the flavor state.
We have
\bea\label{charge2}
{\cal Q}^{\bf k}_{\nu_e \rightarrow\nu_\si}(t) &\equiv&
\langle \nu^r_{{\bf k},e}|
Q_\sigma(t) |\nu^r_{{\bf k},e}\rangle
\,=\, \lf|\lf \{\al^{r}_{{\bf k},\si}(t), \al^{r
\dag}_{{\bf k},e}(0) \ri\}\ri|^{2} \;+ \;\lf|\lf\{\bt_{{-\bf
k},\si}^{r \dag}(t), \al^{r \dag}_{{\bf k},e}(0) \ri\}\ri|^{2}
\eea
and
\bea
 _{e,\mu }\langle 0|\nof\wwQ_{e}(t)\nof|0\rangle _{e,\mu } =
 \, _{e,\mu}\langle 0|Q_{\mu }(t)|0\rangle _{e,\mu }=0,
\eea

The oscillation formulas are \cite{BHV99}:
\begin{eqnarray}  \label{oscillfor1}
{\cal Q}^{\bf k}_{{\nu_e}\rightarrow\nu_e}(t) &=&
1-\sin ^{2}(2\theta )\Big[ \left| U_{\mathbf{k}}\right| ^{2}\sin
^{2}\left( \frac{\omega _{k,2}-\omega _{k,1}}{2}t\right)
+\left| V_{\mathbf{k%
}}\right| ^{2}\sin ^{2}\left( \frac{\omega _{k,2}+\omega
_{k,1}}{2}t\right) \Big],
\\
\label{oscillfor2}
{\cal Q}^{\bf k}_{{\nu_e}\rightarrow\nu_\mu}(t)
&=& \sin ^{2}(2\theta )\Big[ \left|
U_{\mathbf{k}}\right| ^{2}\sin ^{2}\left( \frac{\omega _{k,2}-\omega
_{k,1}}{2}t\right)
+\left| V_{\mathbf{k%
}}\right| ^{2}\sin ^{2}\left( \frac{\omega _{k,2}+\omega
_{k,1}}{2}t\right) \Big].
\end{eqnarray}

The charge conservation is ensured at any time:
\bea
{\cal Q}^{\bf k}_{{\nu_e}\rightarrow\nu_e}(t) +
{\cal Q}^{\bf k}_{{\nu_e}\rightarrow\nu_\mu}(t) = 1.
 \eea

The differences with respect to the Pontecorvo formulas are: the
energy dependence of the amplitudes,
 and the additional oscillating term.
In the relativistic limit: $\left| \mathbf{k}\right| \gg \sqrt{m_{1}m_{2}}$, we have
$\left| U_{\mathbf{k}}\right| ^{2}\longrightarrow 1$ and $\left|
V_{\mathbf{k}}\right| ^{2}\longrightarrow 0$ and the traditional
formulas are recovered.

\section{Entanglement in neutrino mixing --  Quantum Field Theory}

Following what done in the QM case, we now calculate the entanglement
associated to an electron neutrino state at time $t$, by means of the variances of the above discussed charge operators.

Let us start with the Noether charges $Q_{\nu_i}$, which are expected to characterize the amount
of static entanglement present in the states Eq.(\ref{flavstate}). We obtain:
\bea \non
\De Q_{i}(\nu_e)(t) &=&\lan \nu^{r}_{\;{\bf k},e}|Q_{i}^2|\nu^{r}_{\;{\bf k},e}\rangle \,-\,
\lan \nu^{r}_{\;{\bf k},e}| Q_{i}|\nu^{r}_{\;{\bf k},e}\rangle^2
\\ \label{DEQn1}
&=&  \frac{1}{4} \sin^2(2\te)\,, \qquad i=1,2.
\eea
in perfect agreement with the quantum mechanical result Eq.(\ref{statentQM}).

\smallskip

Next we consider  dynamic entanglement, which is described by the variances of the flavor
charges. We have:
\bea \non
\De Q_{e}(\nu_e)(t) &=&\lan \nu^{r}_{\;{\bf k},e}|Q_{e}^2(t)|\nu^{r}_{\;{\bf k},e}\rangle \,-\,
\lan \nu^{r}_{\;{\bf k},e}| Q_{e}(t)|\nu^{r}_{\;{\bf k},e}\rangle^2
\\ \non
&=&  \,\lan \nu^{r}_{\;{\bf k},e}|\lf[ \sum_{s}
\int d^3 {\bf p} \, \lf( \al^{s\dag}_{{\bf p},e}(t)
\al^{s}_{{\bf p},e}(t)\, -\, \beta^{s\dag}_{-{\bf p},e}(t)
\beta^{s}_{-{\bf p},e}(t)\ri)\ri]^2\,|\nu^{r}_{\;{\bf k},e}\rangle
\,-\,
\lf[{\cal Q}^{\bf k}_{{\nu_e}\rightarrow\nu_e}(t)\ri]^2
\\ \non
&=&  \,\lan \nu^{r}_{\;{\bf k},e}| \al^{r\dag}_{{\bf k},e}(t)
\al^{r}_{{\bf k},e}(t)|\nu^{r}_{\;{\bf k},e}\rangle
\,+\,
\lan \nu^{r}_{\;{\bf k},e}| \beta^{r\dag}_{-{\bf k},e}(t)
\beta^{r}_{-{\bf k},e}(t)|\nu^{r}_{\;{\bf k},e}\rangle
\\  \label{DEQ2}
&-& 2 \,
\lan \nu^{r}_{\;{\bf k},e}| \al^{r\dag}_{{\bf k},e}(t)
\al^{r}_{{\bf k},e}(t)\beta^{r\dag}_{-{\bf k},e}(t)
\beta^{r}_{-{\bf k},e}(t)|\nu^{r}_{\;{\bf k},e}\rangle
\,-\,
\lf[{\cal Q}^{\bf k}_{{\nu_e}\rightarrow\nu_e}(t)\ri]^2.
\eea

We now consider the third term in Eq.(\ref{DEQ2}). We have:
\bea \non
\lan \nu^{r}_{\;{\bf k},e}| \al^{r\dag}_{{\bf k},e}(t)
\al^{r}_{{\bf k},e}(t)\beta^{r\dag}_{-{\bf k},e}(t)
\beta^{r}_{-{\bf k},e}(t)|\nu^{r}_{\;{\bf k},e}\rangle
=
\, _{e,\mu }\lan 0| \al^{r\dag}_{{\bf k},e}(t)
\al^{r}_{{\bf k},e}(t)\beta^{r\dag}_{-{\bf k},e}(t)
\beta^{r}_{-{\bf k},e}(t)|0\rangle_{e,\mu }
\\ \non
+\lf|\lf \{\al^{r}_{{\bf k},e}(t), \al^{r\dag}_{{\bf k},e}(0) \ri\}\ri|^{2}
\, _{e,\mu }\langle 0| \beta^{r\dag}_{-{\bf k},e}(t)
\beta^{r}_{-{\bf k},e}(t)|0\rangle _{e,\mu }
\\ \non
- \lf|\lf \{\al^{r}_{{\bf k},e}(0), \bt^{r}_{-{\bf k},e}(t) \ri\}\ri|^{2}
\, _{e,\mu }\langle 0| \al^{r\dag}_{{\bf k},e}(t)
\al^{r}_{{\bf k},e}(t)|0\rangle _{e,\mu }
\\ \non
- \lf \{\al^{r\dag}_{{\bf k},e}(0), \bt^{r\dag}_{-{\bf k},e}(t) \ri\}
\lf \{\al^{r}_{{\bf k},e}(0), \al^{r\dag}_{{\bf k},e}(t) \ri\}
\, _{e,\mu }\langle 0| \al^{r}_{{\bf k},e}(t)
\bt^{r}_{-{\bf k},e}(t)|0\rangle _{e,\mu }
\\
+ \lf \{\al^{r}_{{\bf k},e}(0), \bt^{r}_{-{\bf k},e}(t) \ri\}
\lf \{\al^{r}_{{\bf k},e}(t), \al^{r\dag}_{{\bf k},e}(0) \ri\}
\, _{e,\mu }\langle 0| \al^{r\dag}_{{\bf k},e}(t)
\bt^{r\dag}_{-{\bf k},e}(t)|0\rangle _{e,\mu }
\eea

\begin{figure}[t]
\centering
\includegraphics*[width=7.5cm]{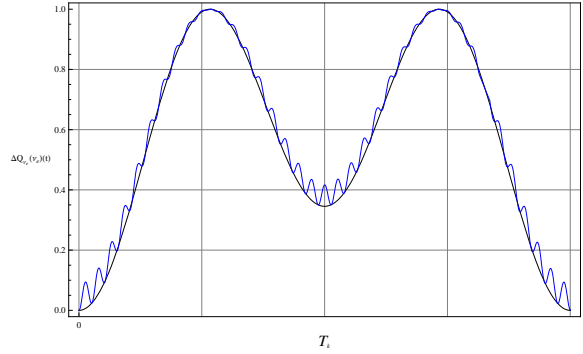}
\caption{QM vs. QFT flavor entanglement
for $|\nu_e(t)\ran$ as a function of the scaled time $T=\frac{2Et}{\Delta m_{12}^{2}}$ with
$\theta$  fixed at the
value $\sin^{2}\theta = 0.314$.
}
\label{FigTwoFlavQFT}
\end{figure}

Explicit calculation of the above quantity shows that:
\bea  \label{DEQ2flavors}
\lan \nu^{r}_{\;{\bf k},e}| \al^{r\dag}_{{\bf k},e}(t)
\al^{r}_{{\bf k},e}(t)\beta^{r\dag}_{-{\bf k},e}(t)
\beta^{r}_{-{\bf k},e}(t)|\nu^{r}_{\;{\bf k},e}\rangle
&=&
\lan \nu^{r}_{\;{\bf k},e}| \beta^{r\dag}_{-{\bf k},e}(t)
\beta^{r}_{-{\bf k},e}(t)|\nu^{r}_{\;{\bf k},e}\rangle
\eea
so that we have
\bea
\De Q_{e}(\nu_e)(t) &=&
{\cal Q}^{\bf k}_{{\nu_e}\rightarrow\nu_e}(t)
\, {\cal Q}^{\bf k}_{{\nu_e}\rightarrow\nu_\mu}(t)
\eea
which formally resembles the quantum  mechanical result Eq.(\ref{DeNQM}).
The differences are now due to the
presence of the flavor condensate, which affect the oscillation formulas (see Eqs.(\ref{oscillfor1}),(\ref{oscillfor2})).
In Fig.\ref{FigTwoFlavQFT}, flavor entanglement formula is plotted in
the QFT case against the corresponding QM case.

\section{Conclusions}

On the basis of recent results, we have discussed some aspect of  entanglement
in the phenomenon of neutrino mixing and  oscillations.
In the  simple case of two flavor mixing, we have shown how to generalize previous
results obtained in the context of quantum mechanics, to the case of quantum field theory.
The difference between the QFT and the QM cases are related to the condensate vacuum structure associated to neutrino mixing in QFT.

Our study provides a simple, exactly solvable example for a possible extension of entanglement to
quantum field theory.
Apart from the differences with QM due to the flavor vacuum contributions, the QFT result is interesting from a more conceptual point of view. Indeed, both the static and the dynamical entanglement arise in connection with unitarily inequivalent representations:
in the case of the static entanglement, the flavor Hilbert space at time $t$ to which the entangled state $|\nu_\si(t)\ran$ belongs, is unitarily inequivalent to the Hilbert space for the qubit states $|\nu_i\ran$ \cite{BV95}; on the other hand, in the case of  dynamical entanglement, where the qubits are taken to be  the flavor states at time $t=0$, the inequivalence is among the flavor Hilbert space at different times \cite{Blasone:2005ae}. In the first case, the relevant orthogonality relation is
$\lim_{V\rar \infty}{}_m\lan 0 |0(t) \rangle_{f} =0$, in the second  $\lim_{V\rar \infty}{}_f\lan 0(t') |0(t) \rangle_{f} =0$, with $t\neq t'$.

Since the inequivalent representations are associated with a non-trivial condensate vacuum structure, the above conjecture suggests that, in the context of QFT, many interpretational issues connected with entanglement could be revisited in this new light.

\appendix

\section{QFT formalism for mixed fields}

The fields
$\nu_{e}(x)$ and $\nu_{\mu}(x)$ are defined through the mixing
relations (\ref{fieldmix}), in terms of the free fields $\nu_{1}(x)$ and $\nu_{2}(x)$
which are expanded as
\bea\label{freefi}
 \nu _{i}(x)=\frac{1}{\sqrt{V}}{\sum_{{\bf k} ,
r}} \left[ u^{r}_{{\bf k},i}\, \al^{r}_{{\bf k},i}(t) +
v^{r}_{-{\bf k},i}\, \bt^{r\dag}_{-{\bf k},i}(t) \ri] e^{i {\bf
k}\cdot{\bf x}},\qquad \qquad i=1,2
\eea
 with $ \al_{{\bf
k},i}^{r}(t)=\al_{{\bf k},i}^{r}\, e^{-i\omega _{k,i}t}$, $
\bt_{{\bf k},i}^{r\dag}(t) = \bt_{{\bf k},i}^{r\dag}\,
e^{i\omega_{k,i}t},$ and $ \omega _{k,i}=\sqrt{{\bf k}^{2} +
m_{i}^{2}}.$ The operator $\alpha ^{r}_{{\bf k},i}$ and $ \beta
^{r }_{{\bf k},i}$, $ i=1,2 \;, \;r=1,2$ are the annihilator
operators for the vacuum state
$|0\rangle_{m}\equiv|0\rangle_{1}\otimes |0\rangle_{2}$: $\alpha
^{r}_{{\bf k},i}|0\rangle_{m}= \beta ^{r }_{{\bf
k},i}|0\rangle_{m}=0$.
 The anticommutation relations are:
$\left\{ \nu _{i}^{\alpha }(x),\nu _{j}^{\beta \dagger
}(y)\right\} _{t=t^{\prime }}=\delta ^{3}({\bf x-y})\delta
_{\alpha \beta } \delta _{ij},$ with $\alpha ,\beta = 1,...4,$
and
$\left\{ \alpha _{{\bf k},i}^{r},\alpha _{{\bf q},j}^{s\dagger
}\right\} =\delta _{{\bf kq}}\delta _{rs}\delta _{ij};$ $\left\{
\beta _{{\bf k},{i}}^{r},\beta _{{\bf q,}{j}}^{s\dagger }\right\}
=\delta _{{\bf kq}}\delta _{rs}\delta _{ij},$ with $i,j=1,2.$ All
other anticommutators vanish. The orthonormality and
completeness relations are given by $u_{{\bf k},i}^{r\dagger }u_{{\bf
k},i}^{s} = v_{{\bf k},i}^{r\dagger }v_{{\bf k},i}^{s} = \delta
_{rs},\; $ $u_{{\bf k},i}^{r\dagger }v_{-{\bf k},i}^{s} = v_{-{\bf
k} ,i}^{r\dagger }u_{{\bf k},i}^{s} = 0,\;$ and $\sum_{r}(u_{{\bf
k},i}^{r}u_{{\bf k},i}^{r\dagger }+v_{-{\bf k},i}^{r}v_{-{\bf
k},i}^{r\dagger }) = \ide.$

The generator of the mixing transformations
 is given by  \cite{BV95}:
\bea\label{generator12}
 G_{\bf \te}(t) = \exp\left[\theta \int
d^{3}{\bf x} \left(\nu_{1}^{\dag}(x) \nu_{2}(x) -
\nu_{2}^{\dag}(x) \nu_{1}(x) \right)\right]\;
\eea
so that
\bea
&&\nu_{\sigma}^{\alpha}(x) = G^{-1}_{\bf
\te}(t)\;
\nu_{i}^{\alpha}(x)\; G_{\bf \te}(t) \,;\qquad
(\sigma,i)=(e,1),(\mu,2)
\eea

At finite volume, this is a unitary operator, $G^{-1}_{\bf
\te}(t)=G_{\bf -\te}(t)=G^{\dag}_{\bf \te}(t)$, preserving the
canonical anticommutation relations. The generator $G^{-1}_{\bf
\te}(t)$ maps the Hilbert space for free fields ${\cal H}_{1,2}$ to
the Hilbert space for mixed fields ${\cal H}_{e,\mu}$: $
G^{-1}_{\bf \te}(t): {\cal H}_{1,2} \mapsto {\cal H}_{e,\mu}.$ In
particular, the flavor vacuum is given by
$
 |0(t) \rangle_{e,\mu} = G^{-1}_{\bf \te}(t)\;
|0 \rangle_{1,2}\; $ at finite
volume $V$. We denote by
$|0 \rangle_{e,\mu}$  the flavor vacuum  at $t=0$.
In the infinite volume limit, the flavor  and the mass vacua
are unitarily
inequivalent \cite{BV95}.
The explicit expression for
$|0\rangle_{e,\mu}$  at time $t=0$ in the reference frame for which
${\bf k}=(0,0,|{\bf k}|)$ is

\bea\label{0emu}
|0\rangle_{e,\mu}^{{\bf k}}&=& \prod_{r}
\Big[(1-\sin^{2}\theta\;|V_{{\bf k}}|^{2})
-\epsilon^{r}\sin\theta\;\cos\theta\; |V_{{\bf k}}|
(\alpha^{r\dag}_{{\bf k},1}\beta^{r\dag}_{-{\bf k},2}+
\alpha^{r\dag}_{{\bf k},2}\beta^{r\dag}_{-{\bf k},1})+
\\ \non
 &+&\epsilon^{r}\sin^{2}\theta \;|V_{{\bf k}}||U_{{\bf
k}}|(\alpha^{r\dag}_{{\bf k},1}\beta^{r\dag}_{-{\bf k},1}-
\alpha^{r\dag}_{{\bf k},2}\beta^{r\dag}_{-{\bf k},2})
+\sin^{2}\theta \; |V_{{\bf k}}|^{2}\alpha^{r\dag}_{{\bf
k},1}\beta^{r\dag}_{-{\bf k},2} \alpha^{r\dag}_{{\bf
k},2}\beta^{r\dag}_{-{\bf k},1} \Big]|0\rangle_{1,2}
 \eea

The condensation density is given by
\bea _{e,\mu}\langle 0| \al_{{\bf k},i}^{r \dag} \al^r_{{\bf k},i}
|0\rangle_{e,\mu}\,= \;_{e,\mu}\langle 0| \bt_{{\bf k},i}^{r \dag}
\bt^r_{{\bf k},i} |0\rangle_{e,\mu}\,=\, \sin^{2}\te\; |V_{{\bf
k}}|^{2} \;, \qquad i=1,2\,. \eea

The flavor fields are written as:
\begin{eqnarray}\label{flavorfield}
\nu _{\sigma}({\bf x},t) &=&\frac{1}{\sqrt{V}}{\sum_{{\bf k},r} }
e^{i{\bf k.x}}\left[ u_{{\bf k},i}^{r}\,  \alpha _{{\bf
k},\sigma}^{r}(t) + v_{-{\bf k},i}^{r} \, \beta _{-{\bf
k},\sigma}^{r\dagger }(t)\right], \quad (\sigma,i)=(e,1),(\mu,2).
\end{eqnarray}
 The flavor annihilation operators are  \cite{BV95}:
\bea \non \alpha^{r}_{{\bf k},e}(t)&=&\cos\theta\;\alpha^{r}_{{\bf
k},1}(t)\;+\;\sin\theta\;\sum_{s}\left[u^{r\dag}_{{\bf k},1}
u^{s}_{{\bf k},2}\; \alpha^{s}_{{\bf k},2}(t)\;+\; u^{r\dag}_{{\bf
k},1} v^{s}_{-{\bf k},2}\; \beta^{s\dag}_{-{\bf k},2}(t)\right]
\\ \non
\alpha^{r}_{{\bf k},\mu}(t)&=&\cos\theta\;\alpha^{r}_{{\bf
k},2}(t)\;- \;\sin\theta\;\sum_{s}\left[u^{r\dag}_{{\bf k},2}
u^{s}_{{\bf k},1}\; \alpha^{s}_{{\bf k},1}(t)\;+\; u^{r\dag}_{{\bf
k},2} v^{s}_{-{\bf k},1}\; \beta^{s\dag}_{-{\bf k},1}(t)\right]
\\ \non
\beta^{r}_{-{\bf k},e}(t)&=&\cos\theta\;\beta^{r}_{-{\bf k},1}(t)\;+
\;\sin\theta\;\sum_{s}\left[v^{s\dag}_{-{\bf k},2} v^{r}_{-{\bf
k},1}\; \beta^{s}_{-{\bf k},2}(t)\;+\; u^{s\dag}_{{\bf k},2}
v^{r}_{-{\bf k},1}\; \alpha^{s\dag}_{{\bf k},2}(t)\right]
\\
\beta^{r}_{-{\bf k},\mu}(t)&=&\cos\theta\;\beta^{r}_{-{\bf
k},2}(t)\;- \;\sin\theta\;\sum_{s}\left[v^{s\dag}_{-{\bf k},1}
v^{r}_{-{\bf k},2}\; \beta^{s}_{-{\bf k},1}(t)\;+\;
u^{s\dag}_{{\bf k},1} v^{r}_{-{\bf k},2}\;
\alpha^{s\dag}_{{\bf k},1}(t)\right].
\label{annih1}
\eea
In the reference frame where ${\bf k}=(0,0,|{\bf k}|)$,
we have
\bea\non
\alpha^{r}_{{\bf k},e}(t)&=&\cos\theta\;\alpha^{r}_{{\bf
k},1}(t)\;+\;\sin\theta\;\left( |U_{{\bf k}}|\; \alpha^{r}_{{\bf
k},2}(t)\;+\;\epsilon^{r}\; |V_{{\bf k}}|\; \beta^{r\dag}_{-{\bf
k},2}(t)\right),
\\ \non
\alpha^{r}_{{\bf k},\mu}(t)&=&\cos\theta\;\alpha^{r}_{{\bf
k},2}(t)\;-\;\sin\theta\;\left( |U_{{\bf k}}|\; \alpha^{r}_{{\bf
k},1}(t)\;-\;\epsilon^{r}\; |V_{{\bf k}}|\; \beta^{r\dag}_{-{\bf
k},1}(t)\right),
\\ \non
\beta^{r}_{-{\bf k},e}(t)&=&\cos\theta\;\beta^{r}_{-{\bf
k},1}(t)\;+\;\sin\theta\;\left( |U_{{\bf k}}|\; \beta^{r}_{-{\bf
k},2}(t)\;-\;\epsilon^{r}\; |V_{{\bf k}}|\; \alpha^{r\dag}_{{\bf
k},2}(t)\right),
\\ \label{annihilator}
\beta^{r}_{-{\bf k},\mu}(t)&=&\cos\theta\;\beta^{r}_{-{\bf
k},2}(t)\;-\;\sin\theta\;\left( |U_{{\bf k}}|\; \beta^{r}_{-{\bf
k},1}(t)\;+\;\epsilon^{r}\; |V_{{\bf k}}|\; \alpha^{r\dag}_{{\bf
k},1}(t)\right).
\eea
Here $\epsilon^{r}=(-1)^{r}$ and
\bea\label{Uk2} \non
 |U_{{\bf k}}| & \equiv & u^{r\dag}_{{\bf k},i}
u^{r}_{{\bf k},j} = v^{r\dag}_{-{\bf k},i} v^{r}_{-{\bf k},j} \,
=\,\frac{|{\bf k}|^{2} +(\om_{k,1}+m_{1})(\om_{k,2}+m_{2})}{2
\sqrt{\om_{k,1}\om_{k,2}(\om_{k,1}+m_{1})(\om_{k,2}+m_{2})}}\,,  \qquad i,j = 1,2, \quad i \neq j,
\\
\label{Vk2}  |V_{{\bf k}}| & \equiv & \epsilon^{r}\; u^{r\dag}_{{\bf
k},1} v^{r}_{-{\bf k},2} = -\epsilon^{r}\; u^{r\dag}_{{\bf k},2}
v^{r}_{-{\bf k},1}\,
=\,\frac{ (\om_{k,1}+m_{1}) - (\om_{k,2}+m_{2})}{2
\sqrt{\om_{k,1}\om_{k,2}(\om_{k,1}+m_{1})(\om_{k,2}+m_{2})}}\, |{\bf k}| \,,\eea
with
$|U_{{\bf
k}}|^{2}+|V_{{\bf k}}|^{2}=1$.

The explicit expressions for  the flavor
states $|\nu_{{\bf k},e}^{r} \rangle $
 and $|\nu_{{\bf k},\mu}^{r}  \rangle $ at time $t=0$, in the reference
frame for which ${\bf k}=(0,0,|{\bf k}|)$ are

\bea  \label{h1}
|\nu_{{\bf k},e}^{r} \rangle &\equiv& \alpha_{{\bf
k},e}^{r \dag}(0)|0\rangle_{e,\mu}
\\ \non
 &=& \left[
\cos\theta\,\alpha_{{\bf k},1}^{r \dag} + |U_{\bf k}|\;
\sin\theta\;\alpha_{{\bf k},2}^{r \dag} - \epsilon^r \; |V_{\bf
k}| \,\sin\theta \; \alpha_{{\bf k},1}^{r \dag}\alpha_{{\bf
k},2}^{r \dag} \beta_{-{\bf k},1}^{r \dag} \right]
G_{{\bf k},s \neq r}^{-1}(\theta) \prod_{{\bf p}\neq{\bf k}} G_{\bf p}^{-1}(\theta)|0\rangle_{1,2},
\\  \label{h2}
|\nu_{{\bf k},\mu}^{r}  \rangle &\equiv& \alpha_{{\bf k},\mu}^{r
\dag}(0)|0\rangle_{e,\mu}
\\ \non
&=& \left[ \cos\theta\,\alpha_{{\bf
k},2}^{r \dag} - |U_{\bf k}|\; \sin\theta\;\alpha_{{\bf k},1}^{r
\dag} + \epsilon^r \; |V_{\bf k}| \,\sin\theta \; \alpha_{{\bf
k},1}^{r \dag}\alpha_{{\bf k},2}^{r \dag} \beta_{-{\bf k},2}^{r
\dag} \right]
G_{{\bf k},s \neq r}^{-1}(\theta) \prod_{{\bf p}\neq{\bf k}} G_{\bf p}^{-1}(\theta)|0\rangle_{1,2}
\,, \eea
where $G (\theta,t) = \prod_{\bf p} \prod_{s = 1}^{2}  G_{{\bf p},s} (\theta,t)$.
In these states a multiparticle component is present, disappearing
in the relativistic limit $|{\bf k}|\gg \sqrt{m_1m_2}\,$: in this
limit, since $\left| U_{\mathbf{k}}\right| ^{2}\longrightarrow 1$ and $\left|
V_{\mathbf{k}}\right| ^{2}\longrightarrow 0$,
the (quantum-mechanical) Pontecorvo states are recovered.

\ack We  acknowledge partial financial support from MIUR, INFN, INFM and CNISM.
M.B. thanks the  organizers of the ``Symmetries in Science Symposium - Bregenz 2009'', for
the very nice and creative atmosphere in which the workshop was held.

\end{document}